\documentclass[aps,prl,twocolumn,groupedaddress,amsmath,amssymb]{revtex4}
\usepackage{graphicx}  
\usepackage{dcolumn}   
\usepackage{bm}        
\usepackage{verbatim}   
\usepackage{color,soul}
\begin{document}

\title{Zeeman-insensitive cooling of a single atom to its two-dimensional motional ground state in tightly focused optical tweezers}
\author{P.~Sompet}
\affiliation{
   The Dodd-Walls Centre for Photonic and Quantum Technologies, Department of Physics, University of Otago, Dunedin, New Zealand   
   }
\author{Y.~H.~Fung}
\affiliation{
   The Dodd-Walls Centre for Photonic and Quantum Technologies, Department of Physics, University of Otago, Dunedin, New Zealand   
   }
   
\author{E.~Schwartz}
\affiliation{
   The Dodd-Walls Centre for Photonic and Quantum Technologies, Department of Physics, University of Otago, Dunedin, New Zealand   
   }
   
\author{M.~D.~J.~Hunter}
\affiliation{
   The Dodd-Walls Centre for Photonic and Quantum Technologies, Department of Physics, University of Otago, Dunedin, New Zealand   
   }

\author{J.~Phrompao}
\affiliation{
   The Dodd-Walls Centre for Photonic and Quantum Technologies, Department of Physics, University of Otago, Dunedin, New Zealand   
   }
   
\author{M.~F.~Andersen}\email{mikkel.andersen@otago.ac.nz}
\affiliation{
   The Dodd-Walls Centre for Photonic and Quantum Technologies, Department of Physics, University of Otago, Dunedin, New Zealand   
   }
   
\date{\today}

\begin{abstract}
We combine near--deterministic preparation of a single atom with Raman sideband cooling, to create a push button mechanism to prepare a single atom in the motional ground state of tightly focused optical tweezers. In the 2D radial plane, we achieve a large ground state fidelity for the entire procedure (loading and cooling) of $\sim$0.73, while the ground state occupancy is $\sim$0.88 for realizations with a single atom present. For 1D axial cooling, we attain a ground state fraction of $\sim$0.52. The combined 3D cooling provides a ground state population of $\sim$0.11. Our Raman sideband cooling variation is indifferent to magnetic field fluctuations, allowing wide--spread unshielded experimental implementations. Our work provides a pathway towards a range of coherent few body experiments.
\end{abstract}

\maketitle


Complete control over individual atoms is vital for gaining a better understanding of the microscopic world as well as enabling new technological pathways. Extensive progress in laser cooled atoms, confined in far off--resonance optical dipole potentials, yields an excellent platform to observe and manipulate matter at the level of single atoms. This has already enabled considerable headway towards quantum logic devices \cite{Saffman1,Saffman2} and quantum simulations \cite{Browaeys2016}, as well as providing detailed insight into microscopic processes whose features are often hidden in ensemble averaged measurements. Such examples are the atomic Hong--Ou--Mandel effect \cite{Regal2014,Greiner2015} and the emergence of statistical mechanics in a quantum state \cite{Greiner2016}. The prospect for further developments in initiating a wide range of effectively zero-entropy quantum states gives this platform unprecedented potential for future studies of few-body physics.

A major challenge in the pursuit of this goal is to prepare atoms in particular quantum states with near--unity fidelity. In cubic geometry, the BEC to Mott--insulator transition allows this for sections of optical lattices \cite{Kuhr2013,Bloch2016}. The flexibility provided by sets of optical tweezers beams \cite{Lukin2016,Rauschenbeutel} makes single atoms in such an ideal building block for diverse few--atom quantum states. A number of avenues are being pursued for high fidelity preparation of a single atom in a particular quantum state. A controlled spill process, utilizing Pauli's exclusion principle, allowed for the isolation of small sets of Fermions from a degenerate sample \cite{Jochim,Moritz}. Separating individual Bosonic helium atoms using penning ionization, prepared individual atoms in the 2D radial ground state of optical tweezers with a fidelity of about 0.5. This was primarily limited by the 50$\%$ chance of ending with no atoms in the tweezers \cite{Truscott}. An alternative approach to achieving a single atom in the vibrational ground state of optical tweezers is first to load the atom and subsequently cool it to its 2D radial \cite{Lukin2013} or 3D \cite{Regal2012} ground state. 

In this paper, we present a push button method to provide a single $^{85}$Rb atom in the motional ground state of an optical trap. The method combines near--deterministic preparation of single atoms \cite{AndersenNPAndersen2013,Regal2015}, with Raman sideband cooling \cite{JessenChuWeiss}. We achieve a record fidelity of $\sim$0.73 for bosons in the 2D motional ground state of the optical tweezers. Our scheme is the first demonstration of Raman sideband cooling of neutral atoms using a Zeeman--insensitive transition and we show it works efficiently despite the high number of photon scattering events required for optical pumping between the relevant internal states. A single pair of Raman beams simultaneously cools both radial dimensions. We obtain the results in an environment where magnetic field fluctuations would be detrimental to previously demonstrated Raman sideband cooling schemes \cite{Lukin2013, Regal2012}. After cooling, the atom is in a state where the long internal state coherence time, achievable with magnetically insensitive transitions \cite{Chu1995,Meschede2005}, can be directly harnessed. Finally, we map the parameters and limitations of the scheme and show that it can be extended to 3D quantum ground state cooling.

\begin{figure}[!t]
	\begin{center}
		\includegraphics[width=0.48\textwidth]{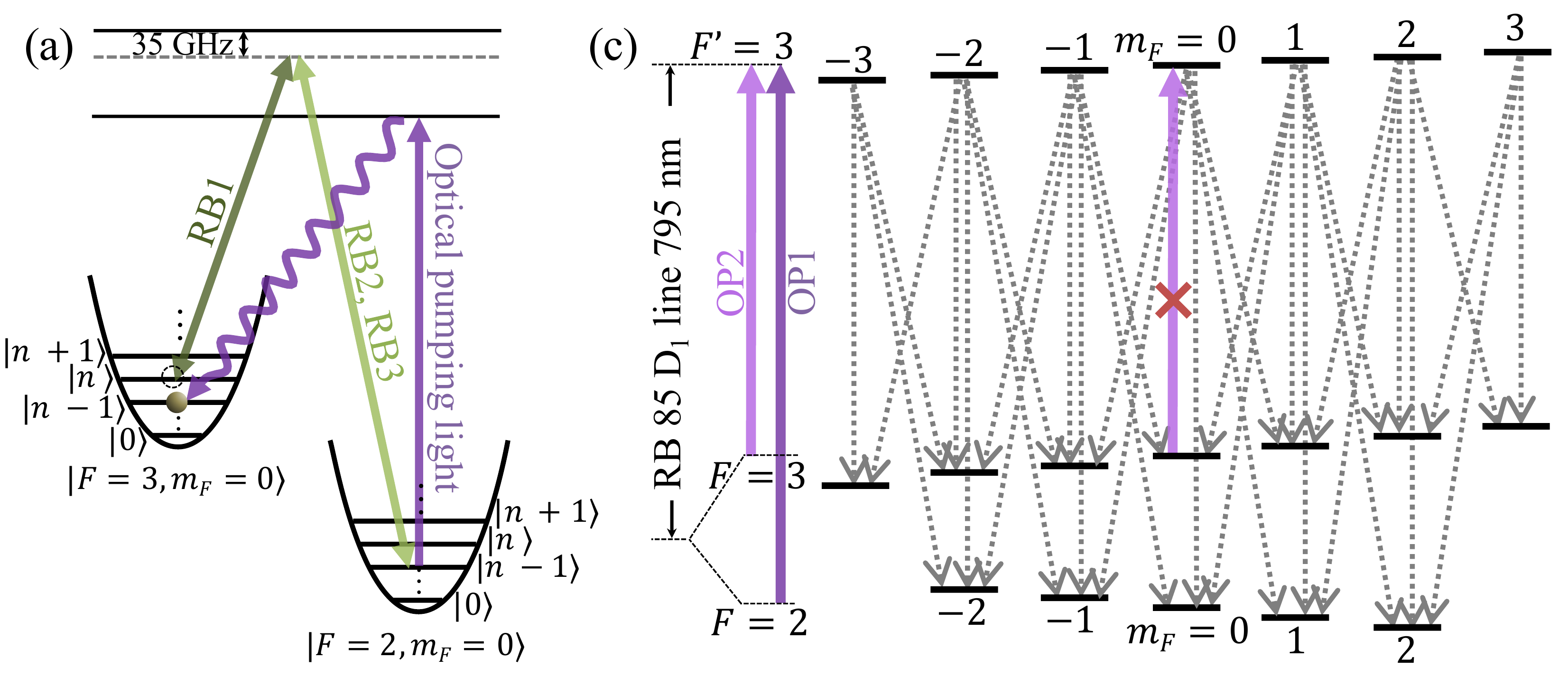}
	\end{center}
	\vspace{-6mm}
	\begin{center}
		\includegraphics[width=0.48\textwidth]{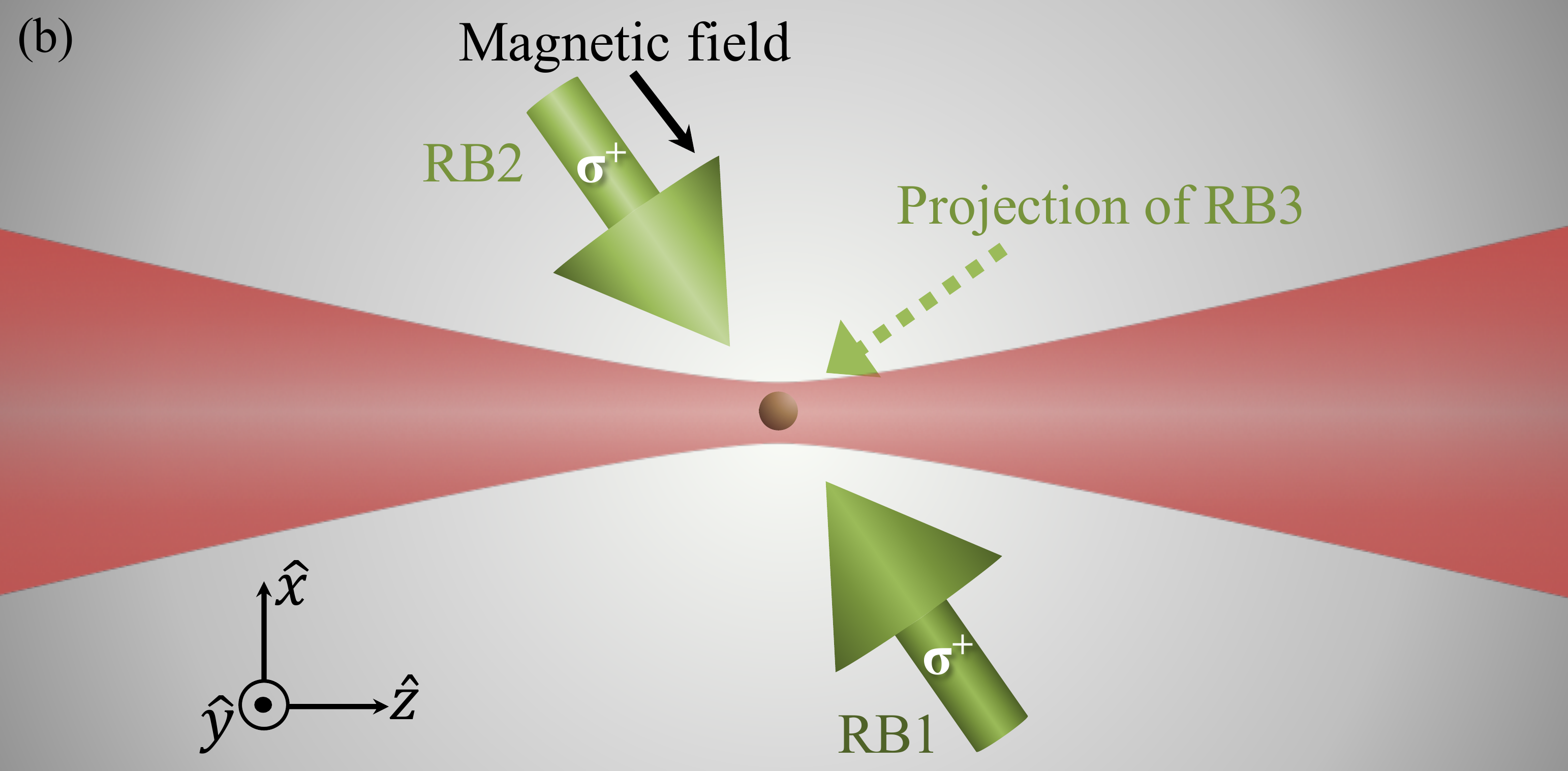}
	\end{center}
	\vspace{-5mm}
   	\caption{(color online). (a) Energy level diagram showing transitions relevant to Raman sideband cooling. (b) Top view of the propagation of Raman beams and optical pumping beams relative to an atom trapped in the optical tweezers. (c) Optical pumping transitions where $|F=3,m_{F}=0\rangle$ is a dark state.}
	\label{fig:figure1}
\end{figure}

Raman sideband cooling efficiently prepares an atom in its vibrational ground state by decoupling the atom from the cooling light once it reaches this state. Figure \ref{fig:figure1}(a) illustrates our utilization of this cooling process. The atom is initially prepared in the $|F,m_{F}\rangle\equiv|3,0\rangle$ internal ground state while being in the $|n\rangle$ state of a harmonic potential with oscillation frequency $\omega$. A pair of Raman beams are tuned to the stimulated Raman transition to the $|2,0\rangle$ internal ground state while stepping down the vibrational state to $|n-1\rangle$. The atom is then optically pumped back to the original internal state ($|3,0\rangle$), thus lowering the energy of the atom if the vibrational state remains $|n-1\rangle$. The entropy of the trapped atom is reduced as the spontaneous emission of an optical pumping photon carries it away. The process will continue until the atom occupies the $|n=0\rangle$ level in the $|3,0\rangle$ internal state, where it is dark for both Raman beams and optical pumping light.

The stimulated Raman transition transfers $\hbar \textbf{$\Delta$k}$ of momentum to the atom/trap system, where $\textbf{$\Delta$k}$ is the wave-vector difference of the two Raman beams. The consequent coupling between final ($|m_x,m_y,m_z\rangle$) and initial ($|n_x,n_y,n_z\rangle$) vibrational states in 3D is represented by the Rabi frequency for the transition \cite{Ye2009}:
\begin{equation}
\begin{split}
\Omega_{R}|\langle m_x,m_y,m_z|e^{i\textbf{$\Delta$k} \cdot \hat{\textbf{R}}}|n_x,n_y,n_z\rangle|=\\
\Omega_{R}\prod_{j=x,y,z}|\langle m_j|e^{i\Delta k_j\hat{R}_j}|n_j\rangle|.
\end{split}
\label{eq:1}
\end{equation}
Here, $\Omega_{R}$ is the Raman coupling parameter between the two internal states, and $\hat{\textbf{R}}$ is the position operator. From the Rabi frequency expression, we can change the vibrational states ($n_j$ to $m_j$) for all three dimensions by using only a single pair of Raman beams as long as $\textbf{$\Delta$k}$ has a projection on all of them. 

In Fig.~\ref{fig:figure1}(b), we present the schematics of our Raman cooling experiment. A strong, linearly polarized, far off--resonance, dipole trap beam ($\lambda=1064$ nm, $\omega_0=1.05$ $\mu$m) propagates along the $\hat{z}$ direction, and holds an atom at the focal point. The applied magnetic field (7.5 Gauss in the $-\sqrt{\frac{2}{3}} \hat{x}+\frac{1}{\sqrt{3}}\hat{z}$ direction) defines the quantization axis of the atom in its internal ground state. Its direction is not aligned with the polarization axis ($\hat{x}$) of the trap beam due to geometric constraints in our experiment. We use three beams to drive Raman transitions (denoted RB1, RB2 and RB3 where the beam peak intensities are 0.5, 1.7 and 2.4 $\times10^3$ mW/cm$^2$ respectively). RB1/RB2 propagates antiparallel/parallel to the magnetic field, and both beams are circularly polarized ($\sigma^{+}$). RB3 propagates orthogonally to the magnetic field (along the $-\frac{1}{\sqrt{6}}\hat{x}+\frac{1}{\sqrt{2}}\hat{y}-\frac{1}{\sqrt{3}}\hat{z}$ direction) and has its linear polarization perpendicular to it as well ($\pi_\perp$), hence it can drive a $\sigma^{\pm}$ transition in the frame defined by the magnetic field. The $\textbf{$\Delta$k}$ of the RB1--RB2 pair thereby has a projection on $\hat{x}$ (radial dimension of the trap) and $\hat{z}$ (axial), while  the $\textbf{$\Delta$k}$ of a RB1--RB3 pair has a projection on the $\hat{y}$ and $\hat{x}$ (radial) directions. The optical pumping light, nearly counter propagates with RB3, and is linearly polarized along the quantization axis.

\begin{figure*}[!t]
	\begin{center}
		\includegraphics[width=1\textwidth]{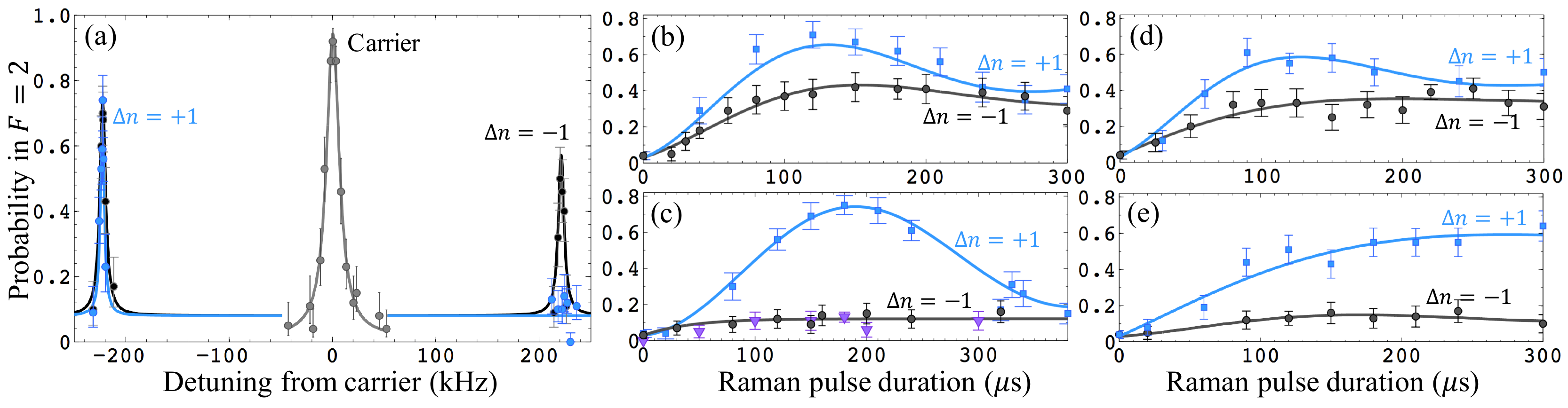}
	\end{center}
	\vspace{-5mm}
   	\caption{(color online). (a) Raman sideband spectrum before (black) and after (blue) the sideband cooling, obtained from spectroscopy by using the RB1--RB3 pair for pulse durations of 90 and 180 $\mu$s respectively. The sideband peaks are fitted with a Lorentzian function, with the solid lines showing the fitted curves. The carrier peak measured using the pulse duration of 40 $\mu$s is also plotted as the grey data set. The offset in the spectrum comes from a combination of the spontaneous emission induced by the Raman beams and the efficiency of the internal state detection. (b)/(c) The transition probability as a function of RB1--RB3 pulse duration at the $\Delta n=-1$ and $\Delta n=+1$ radial sideband peaks before/after the cooling sequence. The transition probability data at an off--resonance (purple triangles) represents the background level. (d)/(e) The transition probability using the RB1--RB2 pair, for before/after the cooling sequence. Data is fitted with damped cosine functions, with the solid lines showing the fitted curves.}
	\label{fig:figure2}
\end{figure*}

Our cooling scheme uses the $|3,0\rangle$ to $|2,0\rangle$ internal state transition which is insensitive to the Zeeman effect to first order. This means that the transition frequency does not change significantly due to the temporal variations in background magnetic fields that prohibit us from using Zeeman--sensitive transitions for Raman sideband cooling. Using the Zeeman--insensitive transition does however have the drawback, that it typically requires a relatively high number of spontaneous photon--scattering events to optically pump the atom back to the initial internal state. This presents a problem in the cooling process, since spontaneous photon--scattering is a source of heating due to the recoil kicks that may change the vibrational quantum number $n$ \cite{Wineland2003}. In sideband cooling schemes this problem is mitigated by the Lamb--Dicke effect that suppresses the probability of changing $n$ for tightly confined atoms \cite{Meekhof1997}. 

Figure \ref{fig:figure1}(c) illustrates our optical pumping light which incorporates two light frequencies matched to the D1 line, denoted OP1 (resonant with the $F=2$ to $F'=3$ transition) and OP2 (resonant with the $F=3$ to $F'=3$ transition). The $\pi$-polarized optical pumping light, cause the atoms to accumulate in the $|3,0\rangle$ internal state given that the transition from this state to the $|3',0\rangle$ excited state is forbidden according to selection rules. After optical pumping, we measure the population of the $|3,0\rangle$ state to be $\sim$0.99 \cite{OPeff}. We use the D1 line for optical pumping because the light shifts from the linearly polarized trap, on both the ground and excited states, are $m_F$ independent. Therefore, the magnetic field defines the quantization axis for optical pumping, even when it is not aligned with the polarization axis of the trap light. Hence, the $|3,0\rangle$ state remains dark in the presence of the deep optical trap. Since the transfer to the $|3,0\rangle$ ground state relies on random changes of $m_F$ and $F$ in the ground state manifold (see Fig. \ref{fig:figure1}(c)), it takes an average of $\sim$9.5 photon scattering events for an atom to transfer from the $|2,0\rangle$ state under ideal conditions. This is significantly higher than the few events required when one uses the maximal $m_F$ states, as is conventionally done \cite{Regal2012, Lukin2013}. The high number of photon--scattering events deteriorates the Raman sideband cooling process if an atom leaves the $|3,0\rangle$ state for reasons other than undergoing the desired stimulated Raman transitions. Moreover, polarization pollution and off--resonant scattering from other excited states dictate that the $|3,0\rangle$ state will not be completely dark to the OP2 light. Therefore, during the cooling cycles we intermittently apply several Raman pulses separated only by OP1 light (OP1 depletes the population in $|2,0\rangle$ state) between every optical pumping pulse that contains both OP1 and OP2 frequencies. This enhances the probability that an atom undergoes a desired Raman transition while suppressing the likelihood of leaving the $|3,0\rangle$ state due to the aforementioned imperfections.      

We start our experimental sequence by laser cooling and preparing a single atom in a tight optical trap using the near--deterministic loading scheme based on engineered blue--detuned light--assisted collisions \cite{AndersenNPAndersen2013}. In our present configuration, the procedure delivers a single atom with a probability of 83$\%$ into a trap with $h\times 57$ MHz depth. We confirm the presence of the atom using fluorescence detection \cite{Andersen2015_img}. To cool the atom to sub--doppler temperatures, we reconfigure the frequency and power of the preparation laser cooling beams for cooling in the deep optical trap. The trap depth is then ramped to $h\times 175$ MHz leaving the single atom with a temperature of 33 $\mu$K (measured by the release-and-recapture (RR) technique \cite{Grangier2008}). At this stage, the trap frequencies are $\{\omega_{x},\omega_{y},\omega_{z}\}/2\pi\simeq\{225,225,36\}$~kHz. Soon after, an optical pumping pulse prepares the atom in the $|3,0\rangle$ state.  After Raman pulses, we determine the population transfer to the $|2,0\rangle$ state by a push--out technique \cite{Meschede2005} that allows us to distinguish the populations of the $F=2$ and $F=3$ ground states (with efficiency of 0.96 for both).

We further characterize the temperature of the atom and the ground state population using sideband spectroscopy. Figure \ref{fig:figure2}(a) shows the Raman spectrum obtained using the RB1--RB3 pair after the initial preparation of the atom. The asymmetry between the height of the $\Delta n=-1$ and $\Delta n=+1$ sideband peaks (denoted $P_{-1}$  and $P_{+1}$ respectively) characterizes the population of the atoms in $|n=0\rangle$ because this state will not contribute to the $\Delta n=-1$ transition. Therefore the mean vibrational quantum number in a particular dimension is $\bar{n} =\frac{P_{-1}/P_{+1}}{1-P_{-1}/P_{+1}}$ \cite{Wineland2003}. Following this, we determine $\{\bar{n}_{x},\bar{n}_{r'},\bar{n}_{z}\}\simeq\{2.4,3.0,20\}$ ($\hat{r}'=\frac{3}{\sqrt{12}}\hat{x}-\frac{1}{2}\hat{y}$, the direction of $\textbf{$\Delta$k}$ for the RB1--RB3 beam pair) which corresponds to temperatures of $\{31,37,35\}$ $\mu$K consistent with the temperature measured by the RR method. 

In Fig.~\ref{fig:figure2}(a), we also present the Raman spectrum obtained after 48 Raman sideband cooling cycles, using the same RB1--RB3 beam pair. The first 24 cooling cycles consist of three Raman beam pulses (50, 90 and 120 $\mu$s) seperated by OP1 light, while the rest consist of a single pulse (100 $\mu$s). The $\Delta n=-1$ sideband peak has nearly vanished, while the $\Delta n=+1$ peak remains, indicating a large atomic population in the ground state. We chose a Raman detuning corresponding to the $\Delta n=-1$ and $\Delta n=+1$ sidebands and measured the transition probability as a function of duration of the Raman pulse. Figure \ref{fig:figure2}(b)/(c) shows the result before/after the cooling. We see damped oscillations before cooling due to the fact that the Rabi frequency differs depending on the $|n\rangle$ initially populated (see Eq.~\ref{eq:1}). After cooling (2c), the $\Delta n=-1$ sideband has vanished while the $\Delta n=+1$ sideband shows coherent Rabi oscillation, showing that only the $|n=0\rangle$ state has a large population.

Figures~\ref{fig:figure2}(d) and (e) reveal that the RB1--RB3 pair efficiently cools both radial dimensions simultaneously, as also observed in \cite{Regal2012}. The figures display similar data to \ref{fig:figure2}(b) and (c) but obtained with RB1--RB2 beam pair after RB1--RB3 cooling. We see that the cooling also leads to a large sideband asymmetry for the RB1--RB2 pair. In Fig.~\ref{fig:figure2}(e), the oscillations are still highly damped. The damping arises since the Rabi frequency of the radial sideband depends on which axial state is occupied when $\textbf{$\Delta$k}$ has a significant projection onto the axial dimension (as is the case for the RB1--RB2 pair). This axial state dependence can be seen from Eq.~\ref{eq:1} which shows that the Rabi frequency of the radial sideband contains the axial carrier matrix element $\left\langle n_{z} \right| \exp ( i \Delta k_z \hat{R_z} ) \left| n_{z}\right\rangle$. The high number of axial states occupied therefore leads to a large range of different Rabi frequencies and the observed damping in Fig.~\ref{fig:figure2}(e) (recall that the axial dimension is not cooled). This effect was weak in Fig.~\ref{fig:figure2}(c) because the  RB1--RB3 pair couples weakly to the axial dimension. From Figs.~\ref{fig:figure2}(c) and (e), we determine the $\bar{n}$ values by using the data where the $\Delta n=+1$ transition probabilities are maximal. We find $\{\bar{n}_{x},\bar{n}_{r'}\}=\{0.08\pm 0.05, 0.04\pm 0.03\}$ with the corresponding ground state population of $\{0.92\pm 0.04, 0.96\pm 0.03\}$. Such 2D cooling occurs if a trap imperfection breaks the radial symmetry and  $\textbf{$\Delta$k}$ has a projection on both the resulting axes while the resulting frequency difference is below the spectral resolution of the Raman pulses. 

We estimate the 2D radial ground state population from the ground state populations measured by sideband asymmetry using the RB1--RB2 and RB1--RB3 pairs separately. Since the $\textbf{$\Delta$k}$ projections of the two pairs on the radial plane are not parallel, they can transfer all non--ground state populations on the $\Delta n=-1$ transition for both radial axes of the trap. A lower bound on the 2D ground state population is thus $0.92\times0.96=0.88$. However, since the $\textbf{$\Delta$k}$ of the two pairs are non-orthogonal it is likely that the 2D ground state population is higher than that. In fact, we saw that the radial symmetry is broken and the $\textbf{$\Delta$k}$ of the RB1--RB3 pair has significant projections on both radial dimensions, as we can achieve efficient 2D cooling using this beam pair alone. Therefore, the ground state population measured by the RB1--RB3 pair (0.96), represents an upper bound of the 2D population. Similarly, the upper bound from the RB1--RB2 pair is consistent with the RB1--RB3 pair value, within the statistical error.

\begin{figure}[!t]
	\begin{center}
		\includegraphics[width=0.5\textwidth]{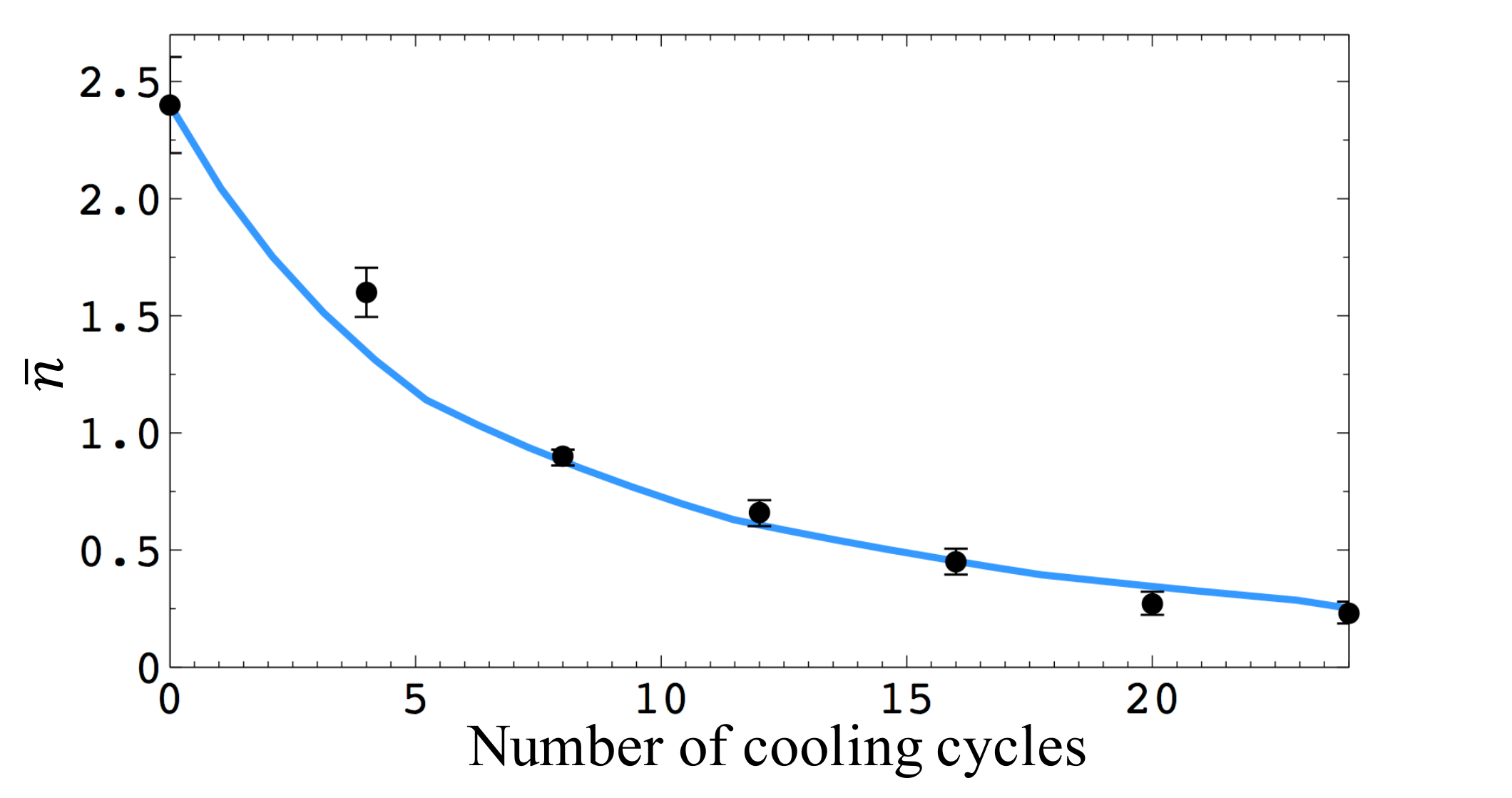}
	\end{center}
	\vspace{-5mm}
   	\caption{(color online). Measured post--cooling $\bar{n}$ as a function of the number of cooling cycles. The blue line is a fit with a simplified model described in the text.}
	\label{fig:figure3}
\end{figure}

In Fig. \ref{fig:figure3}, we characterize the sideband cooling scheme through the evolution of $\bar{n}$ after different number of cooling cycles with the RB1--RB3 pair. The blue line is a fit with a simplified model that assumes the change of energy (in units of $\hbar\omega$) per cooling cycle, $\alpha$, is independent of $n$, except for the ground state, where $\alpha(0)=0$. In this model, we further assume an initial thermal population distribution with temperature, $T$, and calculate the ground state population after a given number of cooling cycles, $c$. Consequently, we get $\bar{n}=[\exp\big(\hbar\omega (\alpha c+1)/(k_\textrm{B}T)\big)-1]^{-1}$ by assuming a Maxwell-Boltzman distribution. The fit gives $\alpha\approx0.15$ which is lower than its ideal value of 0.5 for 2D cooling. Since the Rabi frequency depends on the vibrational excitation number (see Eq.~\ref{eq:1}), when the atomic population occupies different $\left|n\right\rangle$ states, only a portion of the excited state population is transferred in a $\Delta n=-1$ transition with a given Raman pulse duration. This prevents $\alpha$ from reaching the 0.5 bound. Hence, the measured $\alpha$ value indicates that the cooling is efficient despite the high number of photon--scattering events required for optical pumping.

\begin{figure}[!t]
	\begin{center}
		\includegraphics[width=0.45\textwidth]{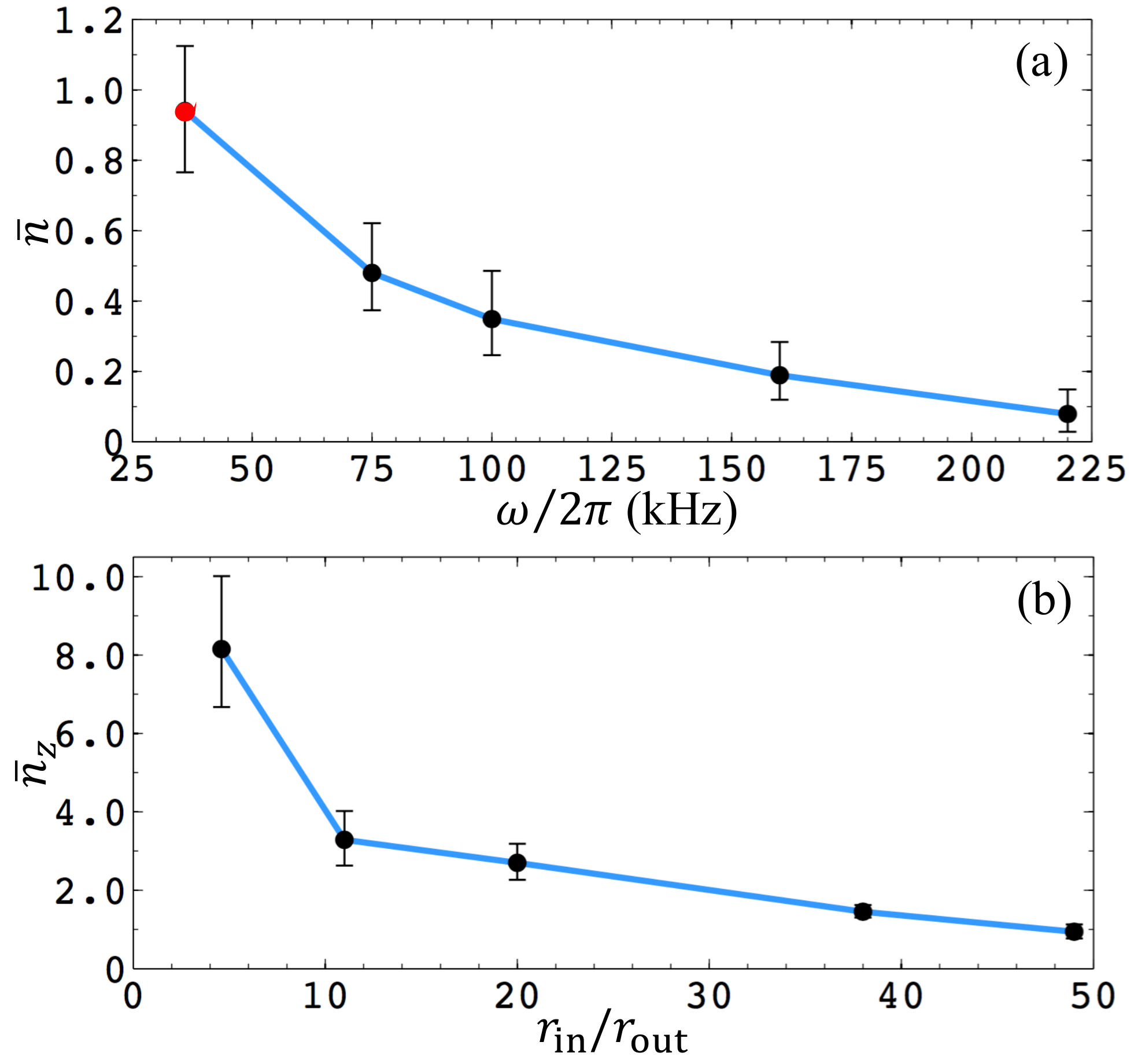}
	\end{center}
	\vspace{-4mm}
   	\caption{(color online). Measured $\bar{n}$ as a function of parameters. (a) $\bar{n}$ as a function of $\omega/(2\pi)$ after 24 cooling cycles with triple Raman pulses. Black points represent cooling on the $r'$ dimension with varied $\omega$ while the red point was obtained by cooling on the $\hat{z}$ axial dimension. (b) $\bar{n}_{z}$ as a function of $r_{\text{in}}/r_{\text{out}}$ ratio after 20 axial cooling cycles.}
	\label{fig:figure4}
\end{figure}

To extend our cooling to 3D, we added axial cooling using the RB1--RB2 beam pair. In our experiments so far, we measure the 3D ground state population to be $\sim$0.11. In our current geometry, the atomic confinement in the axial dimension is relatively weak ($\omega_{z}/2\pi$ is 36 kHz); it follows that the axial motion of the atom is not deep in the Lamb--Dicke regime (Lamb--Dicke parameter of $\eta\approx0.32$), and therefore it is likely that $n_{z}$ changes during the optical pumping stage \cite{Regal2012}. To identify the requirements needed for efficient 3D cooling, we measured $\bar{n}$ after RB1--RB3 beam pair cooling, as we varied the trap frequency. Figure \ref{fig:figure4}(a) presents the results alongside an additional point obtained using axial cooling (in red). Additionally we investigated the effects due to the scattering of optical pumping photons. To quantify the performance of the optical pumping we use a ratio between $r_{\text{in}}$ (the rate of pumping the atoms into the $|3,0\rangle$ state due to OP1 and OP2) and $r_{\text{out}}$ (the rate of pumping the atoms out of the $|3,0\rangle$ state due to OP2). Ideally this ratio should be as large as possible, indicating the least number of photon--scattering events during optical pumping.

We varied the $r_{\text{in}}/r_{\text{out}}$ ratio, by tuning the magnetic field direction, and show the effect of this on $\bar{n}_{z}$ after 20 axial cooling cycles in Fig.~\ref{fig:figure4}(b). Fig.~\ref{fig:figure4} indicates that our 3D ground state population could be significantly enhanced by increasing the axial frequency to surpass $100$ kHz, while a gain from further optimization of the optical pumping would be marginal. In our apparatus, we could access $\omega_{z}/2\pi$ above $100$ kHz by changing the dipole trap wavelength to 850 nm yielding a smaller spot size.

The Zeeman--insensitive ground state cooling works consistently, despite magnetic field fluctuations within the experimental region. These fluctuations cause tens of kHz broadening of magnetically sensitive ground state Raman transitions, which prohibits the use of Zeeman sensitive states. Our Raman sideband cooling variation can therefore be implemented in existing non--shielded experiments. Furthermore, cooling by using the magnetically--insensitive transitions, avoids the internal state decoherence from using a non--paraxial trap beam \cite{Lukin2013} and from motion in spatially varying trap light shifts \cite{Yang}. The high fidelity preparation increases the possibilities for studying few body dynamics. Following that, the fidelity of our system could be further enhanced if we optimize the probability for single atom occupancy before cooling. This can be done by variations of our presently--used near--deterministic loading scheme \cite{Regal2015,AndersenNPAndersen2013}, or through applying atomic sorting \cite{Rauschenbeutel,Regal2015,Lukin2016} to refill the zero occupancies from a reservoir. An alluring option will be to use $^{87}$Rb atoms, which could provide a better cooling efficiency as the atoms have a lower number of internal ground states ($\sim$5.7 photons scattering events required for the optical pumping state).

To conclude, we have accomplished the first demonstration of magnetically--insensitive Raman sideband cooling of neutral atoms, and combined it with the near--deterministic preparation of single atoms. By applying this cooling scheme in an environment with significant magnetic field fluctuations, we achieved efficient cooling in the radial plane by using only one pair of Raman beams. Our cooling scheme variation yielded a 2D ground state population of $\sim$0.88 when a single atom is present. This push button method provides an appreciable fidelity of $\sim$0.73 for single atoms in the 2D radial vibrational ground state of optical tweezers. 

This work was supported by the Marsden Fund Council from Government funding, administered by the Royal Society of New Zealand (Contract number UOO1320). We thank Stephen G. Lipson for his comments on our manuscript.

\newpage \setcounter{footnote}{0}\setcounter{figure}{0}\setcounter{equation}{0}\normalsize

\begin{thebibliography}{00}

\bibitem{Saffman1}
M.~Saffman, 
J.~Phys.~B \textbf{49}, 202001 (2016).

\bibitem{Saffman2}
T.~Xia, M.~Lichtman, K.~M.~Maller, A.~W.~Carr, M.~J.~Piotrowicz, L.~Isenhower, and M.~Saffman, 
Phys.~Rev.~Lett.~\textbf{114}, 100503 (2015).

\bibitem{Browaeys2016}
H.~Labuhn, D.~Barredo, S.~Ravets, S.~de L\'{e}s\'{e}leuc, T.~Macr\'{i}, T.~Lahaye, and A.~Browaeys, 
Nature \textbf{534}, 667-670 (2016).

\bibitem{Regal2014}
A.~M.~Kaufman, B.~J.~Lester, C.~M.~Reynolds, M.~L.~Wall, M.~Foss--Feig, K.~R.~A.~Hazzard, A.~M.~Rey, and C.~A.~Regal, 
Science \textbf{345}, 306-309 (2014).

\bibitem{Greiner2015}
R.~Islam, R.~Ma, P.~M.~Preiss, M.~E.~Tai, A.~Lukin, M.~Rispoli, and M.~Greiner, 
Nature \textbf{528}, 77-83 (2015).

\bibitem{Greiner2016}
A.~M.~Kaufman, M.~Eric Tai, A.~Lukin, M.~Rispoli, R.~Schittko, P.~M.~Preiss, and M.~Greiner, 
Science \textbf{353},794 (2016).

\bibitem{Kuhr2013}
T.~Fukuhara, A.~Kantian, M.~Endres, M.~Cheneau, P.~Schau$\beta$, S.~Hild, D.~Bellem, U.~Schollw\"{o}ck, T.~Giamarchi, C.~Gross, I.~Bloch, and S.~Kuhr, 
Nature Phys.~\textbf{9}, 241 (2013).

\bibitem{Bloch2016}
J.~Choi, S.~Hild, J.~Zeiher, P.~Schau$\beta$, A.~Rubio--Abadal, T.~Yefsah, V.~Khemani, D.~A.~Huse, I.~Bloch, and C.~Gross, 
Science \textbf{352}, 1547-1552 (2016).

\bibitem{Rauschenbeutel}
Y.~Miroshnychenko, W.~Alt, I.~Dotsenko, L.~F\"{o}rster, M.~Khudaverdyan, D.~Meschede, D.~Schrader, and A.~Rauschenbeutel, 
Nature \textbf{442}, 151-151 (2006).

\bibitem{Lukin2016}
M.~Endres, H.~Bernien, A.~Keesling, H.~Levine, E.~R.~Anschuetz, A.~Krajenbrink, C.~Senko, V.~Vuletic, M.~Greiner, and M.~D.~Lukin, 
Science \textbf{354}, 1024-1027 (2016).
D. Barredo, S.~de L\'{e}s\'{e}leuc, V. Lienhard, T.~Lahaye, and A.~Browaeys. 
Science \textbf{354}, 1021-1023 (2016).

\bibitem{Jochim}
F.~Serwane, G.~Z\"{u}rn, T.~Lompe, T.~B.~Ottenstein, A.~N.~Wenz, and S.~Jochim, 
Science \textbf{332}, 336 (2011).

\bibitem{Moritz}
B.~Zimmermann, T.~Mueller, J.~Meineke, T.~Esslinger, and H.~Moritz, 
New J.~Phys.~\textbf{13}, 043007 (2011).

\bibitem{Truscott}
A.~G.~Manning, R.~Khakimov, R.~G.~Dall, and A.~G.~Truscott, 
Phys.~Rev.~Lett.~\textbf{113}, 130403 (2014).

\bibitem{Lukin2013}
J.~D.~Thompson, T.~G.~Tiecke, A.~S.~Zibrov, V.~Vuleti\'{c}, and M.~D.~Lukin, 
Phys.~Rev.~Lett.~\textbf{110}, 133001 (2013). 

\bibitem{Regal2012}
A.~M.~Kaufman, B.~J.~Lester, and C.~A.~Regal, 
Phys.~Rev.~X \textbf{2}, 041014 (2012).

\bibitem{AndersenNPAndersen2013}
T.~Grunzweig,	A.~Hilliard,	M.~McGovern,	and M.~F.~Andersen
Nature Phys.~\textbf{6}, 951 (2010). A.~V.~Carpentier, Y.~H.~Fung, P.~Sompet, A.~J.~Hilliard, T.~G.~Walker, and M.~F.~Andersen, 
Laser Phys.~Lett.~\textbf{10}, 125501 (2013). Y.~H.~Fung, and M.~F.~Andersen, 
New J.~Phys.~\textbf{17}, 073011 (2015).

\bibitem{Regal2015}
B.~J.~Lester, N.~Luick, A.~M.~Kaufman, C.~M.~Reynolds, and C.~A.~Regal, 
Phys.~Rev.~Lett.~\textbf{115}, 073003 (2015).

\bibitem{JessenChuWeiss}
S.~E.~Hamann, D.~L.~Haycock, G.~Klose, P.~H.~Pax, I.~H.~Deutsch, and P.~S.~Jessen, 
Phys.~Rev.~Lett.~\textbf{80}, 4149 (1998). V.~Vuleti\'{c}, C.~Chin, A.~J.~Kerman, and S.~Chu, 
Phys.~Rev.~Lett.~\textbf{81}, 5768 (1998). X.~Li, T.~A.~Corcovilos, Y.~Wang, and D.~S.~Weiss, 
Phys.~Rev.~Lett.~\textbf{108}, 103001 (2012).

\bibitem{Chu1995}
N.~Davidson, H.~J.~Lee, C.~S.~Adams, M.~Kasevich, and S.~Chu, 
Phys.~Rev.~Lett.~\textbf{74}, 1311 (1995).

\bibitem{Meschede2005}
S.~Kuhr, W.~Alt, D.~Schrader, I.~Dotsenko, Y.~Miroshnychenko, A.~Rauschenbeutel, and D.~Meschede, 
Phys.~Rev.~A \textbf{72}, 023406 (2005). 

\bibitem{Ye2009}
S.~Blatt, J.~W.~Thomsen, G.~K.~Campbell, A.~D.~Ludlow, M.~D.~Swallows, M.~J.~Martin, M.~M.~Boyd, and J.~Ye, 
Phys.~Rev.~A \textbf{80}, 052703 (2009).

\bibitem{Wineland2003}
D.~Leibfried, R.~Blatt, C.~Monroe, and D.~J.~Wineland, 
Rev.~Mod.~Phys.~\textbf{75}, 281 (2003).

\bibitem{Meekhof1997}
D.~J.~Wineland, C.~Monroe, W.~M.~Itano, D.~Leibfried, B.~E.~King, and D.~M.~Meekhof, 
arXiv:quant-ph/9710025 (1997).

\bibitem{OPeff}
Measurment of optical pumping efficiency: after optical pumping, we introduce OP1 to deplete the $F=2$ ground state population. Then we transfer all $\left|3,0\right\rangle$ population to the $\left|2,0\right\rangle$ state by using co-propagating Raman beams and measure the $F=2$ state population. 

\bibitem{Andersen2015_img}
A.~J.~Hilliard, Y.~H.~Fung, P.~Sompet, A.~V.~Carpentier, M.~F.~Andersen,
Phys.~Rev.~A  \textbf{91}, 053414 (2015). M.~McGovern, A.~J.~Hilliard, T.~Grunzweig, and M.~F.~Andersen, 
Opt.~Lett.~\textbf{36}, 1041 (2011)

\bibitem{Grangier2008}
C.~Tuchendler, A.~M.~Lance, A.~Browaeys, Y.~R.~P.~Sortais, and P.~Grangier,
Phys.~Rev.~A~\textbf{78}, 033425 (2008).

\bibitem{Kimble2006}
A.~D.~Boozer, A.~Boca, R.~Miller, T.~E.~Northup, and H.~J.~Kimble, 
Phys.~Rev.~Lett.~\textbf{97}, 083602 (2006).

\bibitem{Ye2015}
M.~Yeo, M.~T.~Hummon, A.~L.~Collopy, B.~Yan, B.~Hemmerling, E.~Chae, J.~M.~Doyle, and J.~Ye, 
Phys.~Rev.~Lett.~\textbf{114}, 223003 (2015).

\bibitem{DeMille2016}
E.~B.~Norrgard, D.~J.~McCarron, M.~H.~Steinecker, M.~R.~Tarbutt, and D.~DeMille, 
Phys.~Rev.~Lett.~\textbf{116}, 063004 (2016).

\bibitem{Lahaye2016}
A.~Browaeys, D.~Barredo, and T.~Lahaye, 
J. Phys. B \textbf{49}, 152001 (2016).

\bibitem{Yang}
J.~Yang, X.~He, R.~Guo, P.~Xu, K.~Wang, C.~Sheng, M.~Liu, J.~Wang, A.~Derevianko, and M.~Zhan
Phys.~Rev.~Lett. \textbf{117}, 123201 (2016).


\end{thebibliography}
\end{document}